\documentclass[submission,copyright,creativecommons]{eptcs}
\providecommand{\event}{FROM 2025} 

\usepackage{iftex}
\usepackage{graphicx}
\usepackage{array}
\usepackage[disable]{todonotes}
\usepackage{alltt}
\usepackage{listings}
\usepackage{amsmath}
\usepackage{cleveref}

\usepackage[normalem]{ulem}
\usepackage[T1]{fontenc}        
\usepackage{amsmath,amssymb}
\usepackage{algpseudocode}
\usepackage{pifont}

\usepackage{dirtytalk}

\newcommand{\comment}[1]{}

\newcommand{\defectCategory}[1]{\textsf{#1}}
\newcommand{\keyTerm}[1]{\emph{#1}}
\newcommand{\codeElement}[1]{\texttt{#1}}

\usepackage{listings, xcolor}

\definecolor{verylightgray}{rgb}{.97,.97,.97}

\lstdefinelanguage{Solidity}{
	keywords=[1]{anonymous, assembly, assert, balance, break, call, callcode, case, catch, class, constant, continue, constructor, contract, debugger, default, delegatecall, delete, do, else, emit, event, experimental, export, external, false, finally, for, function, gas, if, implements, import, in, indexed, instanceof, interface, internal, is, length, library, log0, log1, log2, log3, log4, memory, modifier, new, payable, pragma, private, protected, public, pure, push, require, return, returns, revert, selfdestruct, send, solidity, storage, struct, suicide, super, switch, then, this, throw, transfer, true, try, typeof, using, value, view, while, with, addmod, ecrecover, keccak256, mulmod, ripemd160, sha256, sha3}, 
	keywordstyle=[1]\color{blue}\bfseries,
	keywords=[2]{address, bool, byte, bytes, bytes1, bytes2, bytes3, bytes4, bytes5, bytes6, bytes7, bytes8, bytes9, bytes10, bytes11, bytes12, bytes13, bytes14, bytes15, bytes16, bytes17, bytes18, bytes19, bytes20, bytes21, bytes22, bytes23, bytes24, bytes25, bytes26, bytes27, bytes28, bytes29, bytes30, bytes31, bytes32, enum, int, int8, int16, int24, int32, int40, int48, int56, int64, int72, int80, int88, int96, int104, int112, int120, int128, int136, int144, int152, int160, int168, int176, int184, int192, int200, int208, int216, int224, int232, int240, int248, int256, mapping, string, uint, uint8, uint16, uint24, uint32, uint40, uint48, uint56, uint64, uint72, uint80, uint88, uint96, uint104, uint112, uint120, uint128, uint136, uint144, uint152, uint160, uint168, uint176, uint184, uint192, uint200, uint208, uint216, uint224, uint232, uint240, uint248, uint256, var, void, ether, finney, szabo, wei, days, hours, minutes, seconds, weeks, years},	
	keywordstyle=[2]\color{teal}\bfseries,
	keywords=[3]{block, blockhash, coinbase, difficulty, gaslimit, number, timestamp, msg, data, gas, sender, sig, value, now, tx, gasprice, origin},	
	keywordstyle=[3]\color{violet}\bfseries,
	identifierstyle=\color{black},
	sensitive=true,
	comment=[l]{//},
	morecomment=[s]{/*}{*/},
	commentstyle=\color{gray}\ttfamily,
	stringstyle=\color{red}\ttfamily,
	morestring=[b]',
	morestring=[b]"
}

\title{Validating Solidity Code Defects using Symbolic and Concrete Execution powered by Large Language Models}

\author{Ștefan-Claudiu Susan
\institute{Alexandru Ioan Cuza University of Iași,\\ Department of Computer Science\\
Iași, România}
\email{claudiu\_susan@yahoo.com}
\and
Andrei Arusoaie
\institute{Alexandru Ioan Cuza University of Iași,\\ Department of Computer Science\\
Iași, România}
\email{andrei.arusoaie@uaic.ro}  
\and
Dorel Lucanu
\institute{Alexandru Ioan Cuza University of Iași,\\ Department of Computer Science\\
Iași, România}
\email{dorel.lucanu@gmail.com}
}

\def\titlerunning{Proving Solidity Code Defects}
\def\authorrunning{S.C. Susan, A. Arusoaie, and D. Lucanu}

\begin{document}

\maketitle

\begin{abstract}
The high rate of false alarms from static analysis tools and Large Language Models (LLMs) complicates vulnerability detection in Solidity Smart Contracts, demanding methods that can formally or empirically prove the presence of defects. This paper introduces a novel detection pipeline that integrates custom Slither-based detectors, LLMs, Kontrol, and Forge. Our approach is designed to reliably detect defects and generate proofs. 

We currently perform experiments with promising results for seven types of critical defects. We demonstrate the pipeline's efficacy by presenting our findings for three vulnerabilities—Reentrancy, Complex Fallback, and Faulty Access Control Policies—that are challenging for current verification solutions, which often generate false alarms or fail to detect them entirely. We highlight the potential of either symbolic or concrete execution in correctly classifying such code faults. By chaining these instruments, our method effectively validates true positives, significantly reducing the manual verification burden. Although we identify potential limitations, such as the inconsistency and the cost of LLMs, our findings establish a robust framework for combining heuristic analysis with formal verification to achieve more reliable and automated smart contract auditing.

\end{abstract}

\section{Introduction}
\label{sec:introduction}
Since the emergence of blockchain platforms like \keyTerm{Ethereum}~\cite{Ethereum}, developers have implemented numerous Decentralized Applications (DApps) across diverse domains, from gaming to decentralized finance. \keyTerm{Solidity}~\cite{Solidity} remains the most widely adopted programming language for the Ethereum ecosystem. However, like any emerging technology, this development paradigm introduced critical shortcomings. The impact of these defects is magnified by two of blockchain's core pillars: immutability, which historically prevented faulty code from being replaced, and public bytecode, which allows malicious actors to easily search for exploits. The devastating potential of such vulnerabilities was demonstrated by catastrophic events, including the \say{DAO Hack}~\cite{DAO} and the \say{Parity Wallet Hack}~\cite{ParityWallet}, which resulted in hundreds of millions of dollars in losses.

The issues identified in Smart Contracts feature unique categories specific to the Blockchain environment. For instance, prominent examples include \defectCategory{Reentrancy}, a critical vulnerability where an external call allows an attacker's contract to repeatedly re-enter a function before its state has been updated, often leading to the complete draining of the contract's funds. Another distinct category involves \defectCategory{Gas-Costly Patterns}, which are not traditional security flaws but rather inefficient coding practices. These patterns, such as unbounded loops or excessive data storage, result in prohibitively high transaction fees (gas) for users, potentially rendering the contract economically unviable or unusable. In addition to these new types of defects, Solidity still features code faults that are common among most programming languages and software development paradigms, such as \defectCategory{Division by Zero} or \defectCategory{Integer Overflow/Underflow}. A crucial step in addressing smart contract flaws has been the development of vulnerability taxonomies. Prior efforts in this area include both formal academic classifications aimed at creating a comprehensive understanding of bug types~\cite{rameder2022review}, and dynamic, community-led efforts (e.g., \say{SWC Registry}~\cite{SwcRegistry}, \say{DASP Top 10}~\cite{DaspTop10}), which serve as practical guides for developers.

State-of-the-art detection tools for smart contract vulnerabilities remain however limited. They often miss real flaws or raise false alarms. 
The types of vulnerabilities range from coding mistakes to challenges inherent to the execution model of smart contracts inside the blockchain ecosystem. 
For instance, the gas system constrains execution by attaching a cost to every instruction, and insufficient gas may alter the expected behavior of a contract. The transactional nature, where failures revert the entire execution, can be abused to mount various attacks (e.g., denial-of-service). 
In addition, the entities executing transactions, such as miners (in proof of work~\cite{Bitcoin}) or validators (in proof of stake~\cite{buterin2022proof}), may themselves act maliciously.
Therefore, it is difficult for a detection tool to address this diversity of vulnerabilities. 

A simple example of a scenario falsely detected as a possible vulnerability is shown in Listing~\ref{lst:reentrancy_false_alarm}.
This contract allows users to either deposit or withdraw Ether, essentially simulating a bank.
The \codeElement{deposit} function receives Ether and records the balance of the caller in a mapping called \codeElement{balance}.
The withdrawal operation uses \codeElement{transfer()} — a function that sends Ether from a contract to a specified address.

Many detection tools (e.g., Slither~\cite{Slither}) flag this contract as ``reentrant'' because the balance is updated after the call to \codeElement{transfer()}.
More specifically, if the \codeElement{withdraw()} function is called from another contract 
C, the \codeElement{transfer()} call triggers a callback to C. 
This happens because, in Ethereum, whenever a smart contract is called, there is a default function named \codeElement{receive()} that is executed automatically. 
If C is malicious, it can exploit the behavior of this default function to immediately invoke another call to \codeElement{withdraw()}.
As a result, the balance mapping may not be updated in time, allowing the attacker to drain funds from \codeElement{ReentrancySimple} as long as funds remain available.

However, in newer versions of Solidity, the \codeElement{transfer()} function enforces a fixed gas limit of 2300 units.
If this gas is exceeded, the entire transaction is reverted.
Since 2300 gas is insufficient for 
C to reinitiate a second call to \codeElement{withdraw()}, \defectCategory{Reentrancy} is no longer possible.

\begin{lstlisting}[language=Solidity, caption={An example of a Reentrancy vulnerability false alarm}, label={lst:reentrancy_false_alarm}]
// SPDX-License-Identifier: MIT
pragma solidity 0.8.29;

contract ReentrancySimple {
    mapping(address => uint) private balance;

    function deposit() external payable {
        balance[msg.sender] = msg.value;
    }

    function withdraw() external {
        uint addrBal = balance[msg.sender];
        //transfer does not trigger Reentrancy
        payable(msg.sender).transfer(addrBal);
        balance[msg.sender] = 0;
    }
}
\end{lstlisting}

\begin{figure}[ht]
\centering
    \includegraphics[width=1\textwidth]{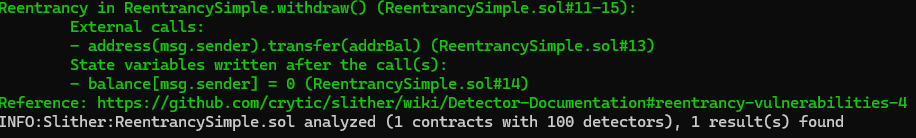} 
    \caption{Output provided by Slither for Listing \ref{lst:reentrancy_false_alarm}}
    \label{fig:Slither_Output}
\end{figure}

During a previous study~\cite{ESE_Paper}, we noticed that Slither flags this example as vulnerable to \defectCategory{Reentrancy} attacks, which is a false alarm. Its exact output can be seen in Figure~\ref{fig:Slither_Output}. The detection mechanisms employed by Slither do not feature concrete or symbolic execution. Being able to perform one of these two techniques on the contract would have shown that a \defectCategory{Reentrancy} attempt always fails. For example, Mythril~\cite{Mythril}, a tool that uses symbolic execution, was able to correctly mark this contract as \say{not vulnerable}. However, it was unable to yield valid detections for other instances of different types of vulnerabilities, where Slither performed considerably better. Thus, we conjecture that a mixture of these two approaches might produce the best overall results.

If we switch the context to the detection capabilities of the large language models, \keyTerm{Google Gemini 2.5 Flash}~\cite{geminiteam2024gemini15unlockingmultimodal} and other popular models were also tricked by this scenario. Exactly as Slither, they signaled that the contract presented in Listing~\ref{lst:reentrancy_false_alarm} was vulnerable to \defectCategory{Reentrancy} exploits. When further questioned, \keyTerm{Gemini 2.5 Flash} falsely stated that 2300 gas is enough to call a function.  Since the popularization of LLMs, multiple studies (see, e.g., \cite{SAS2025,chen2023chatgpt,xiao2025logic}) on their potential for identifying Solidity defects have been conducted. Even though these studies used different datasets, did not use the same AI models, and varied in some methodological aspects, a consistent conclusion emerged: LLMs can detect many issues that traditional tools cannot detect, but their usefulness is limited by false alarms and consistency issues.


This paper introduces a novel, three-stage mechanism to validate potential smart contract vulnerabilities and eliminate false positives. First, we employ a custom static analysis tool built on Slither modules to scan for specific coding patterns that are prerequisites for certain bugs. For example, a Reentrancy analysis is only triggered if the contract contains logic for handling Ether or tokens, as this vulnerability's callbacks typically occur during currency transfers. If these prerequisite patterns are present, our second stage uses a Large Language Model (LLM) to generate a concrete proof-of-exploit. To improve the accuracy of this output, we guide the LLM using bespoke \say{testing patterns} tailored to each defect class. For the third and final stage, we execute the tests using Kontrol~\cite{FormVerificationEVM,Kevm} or Forge~\cite{Forge} (depending on the issue type) to verify the presence of a defect, or its absence within the scope of our test scenarios, through either concrete or symbolic execution. A successfully generated and executed test serves as a formal proof, if Kontrol is used, or as an empirical proof, if Forge is used. The primary limitation of this experimental approach is its dependency on the underlying LLM's capability to accurately follow our structured prompts.
\paragraph{Summary of contributions}
\begin{enumerate}
\setlength\itemsep{0em}
    \item We identify some shortcomings that are present in state-of-the-art defect detection tools, such as not identifying or providing false alarms for the code defects that are discussed in this paper, and present a list of examples that highlight the potential of our approach in 
    Section~\ref{sec:case_study}.
    \item We propose an approach that is custom-made for each of the \textbf{seven} defect types that we have experimented with so far in Section~\ref{sec:our_approach}. The use of the proposed approach is illustrated on an example in Section~\ref{sec:work-ex}
    \item We present the challenges and limitations of the proposed approach in~\Cref{sec:chal-and-lim}.
\end{enumerate}

\paragraph{Paper organization}
In Section~\ref{sec:background_motivation}, we briefly recall information about the blockchain and smart contracts, foundational tools (Forge, Kontrol, Slither), some of our previous work, and related approaches. To validate our proposal, we present multiple case studies in Section~\ref{sec:case_study} that analyze scenarios where other tools fail. We detail our novel detection pipeline in Section~\ref{sec:our_approach}. We provide an example of how our pipeline analyzes a contract to check for a specific defect category in Section~\ref{sec:work-ex}. A discussion of our approach's future potential and the challenges that we might encounter follows in Section~\ref{sec:chal-and-lim}. We conclude in Section~\ref{sec:conclusions}.
\section{Background \& Motivation}
\label{sec:background_motivation}
\subsection{Blockchain \& Smart Contracts}
A blockchain is a decentralized, distributed, and cryptographically secured digital ledger. Its architecture, first conceptualized for a peer-to-peer electronic cash system \cite{Bitcoin}, creates a trustless environment where participants can interact directly without a central intermediary. This is achieved through a combination of core properties: decentralization, which distributes control across a network; transparency, where all participants share a common view of the ledger; and immutability, which ensures that transactions, once recorded in a block and added to the chain, are computationally infeasible to alter.

The capabilities of blockchain technology were significantly expanded with the introduction of smart contracts. The concept, which predates blockchain, describes computer-observable and enforceable protocols that can self-execute the terms of an agreement~\cite{szabo1996smart}. The Ethereum platform was the first to realize this vision on a global scale, providing a permissionless blockchain with a Turing-complete virtual machine known as the Ethereum Virtual Machine (EVM)~\cite{Ethereum,wood2014ethereum}. This innovation enabled the creation of complex Decentralized Applications (DApps) for a wide range of use cases beyond currency transfers.

However, the \say{code is law} paradigm of smart contracts presents a double-edged sword. The principle of immutability, while fundamental to the security and integrity of the ledger, means that bugs and vulnerabilities in deployed contract code are permanent. A flawed contract cannot be easily patched or updated, leaving it open to perpetual exploitation. This high-stakes environment, where simple coding errors can lead to catastrophic financial losses, requires the use of advanced and rigorous analysis techniques to ensure the security and correctness of smart contracts before they are deployed~\cite{SurveyEthereumAttacks}.

\subsection{Tools that Might be Used for the Development and Analysis of Smart Contracts}
\keyTerm{Slither}~\cite{Slither} is a powerful and widely adopted static analysis framework for Solidity. It converts smart contract source code into an intermediate representation called SlithIR, which provides a high-level view of the contract's logic and control flow. Upon this representation, Slither executes a comprehensive suite of vulnerability detectors, which are designed to identify a broad spectrum of issues, ranging from critical security flaws like \defectCategory{Reentrancy} and incorrect access control to code quality and optimization problems. 
Besides these strong points, Slither is an open-source tool that encourages the community to build their tools using the modules that it provides or contribute to improving Slither itself. This allows developers to focus on implementing useful and innovative analysis techniques, building upon the information that is accessible through importing Slither modules into their code. For example, using Slither modules, one can extract the CFG, the parameters of a function, the state variables of a contract, or the variables read/written by each statement. 

\keyTerm{Foundry}~\cite{Foundry} is a modern, high-performance toolkit for Ethereum application development, written entirely in Rust. Unlike traditional JavaScript-based frameworks, Foundry provides a comprehensive and cohesive development environment that allows for faster and more efficient workflows. It consists of several core components: \keyTerm{Forge}, a testing framework that enables developers to write tests directly in Solidity; \keyTerm{Cast}, a versatile command-line tool for performing RPC calls to interact with smart contracts and query blockchain data; \keyTerm{Anvil}, a fast local testnet node for development and debugging; and \keyTerm{Chisel}, an interactive Solidity shell for rapid prototyping. By combining blazing-fast compilation and execution speeds with a Solidity-centric design, Foundry empowers developers to build, test, and deploy smart contracts with greater productivity and confidence. 

\keyTerm{Forge}~\cite{Forge} is renowned for its exceptional speed in compiling contracts and executing tests. It offers a rich set of features designed for robust contract verification, including native fuzz testing to automatically discover edge-case vulnerabilities with random inputs, detailed gas reports to help optimize transaction costs, and a powerful set of \say{cheatcodes} that allow developers to manipulate the blockchain state (e.g., block number, msg.sender) for simulating complex scenarios. This combination of speed, usability, and advanced testing capabilities makes Forge an essential tool for modern smart contract development and auditing.

\keyTerm{Kontrol}~\cite{FormVerificationEVM} is a formal verification tool designed for EVM-based smart contracts and integrated within the Foundry ecosystem. It enables developers to prove the correctness of their code with mathematical certainty, moving beyond the limitations of conventional testing. At its core, Kontrol leverages K-EVM~\cite{Kevm}, a formal and executable specification of the Ethereum Virtual Machine written in the K Framework~\cite{KFramework}. By employing symbolic execution, Kontrol analyzes all possible execution paths of a contract simultaneously. Developers write properties, or \say{lemmas}, as functions directly within their Solidity test files, specifying invariants that must always hold true, such as the absence of \defectCategory{Reentrancy} or the correctness of token accounting. Kontrol then attempts to formally prove that these properties are unbreakable under any circumstance, providing a powerful method for verifying the security and functional correctness of high-value smart contracts.
\subsection{Related Work}
We are building upon our prior efforts~\cite{from2024,from2023}, in which we enhanced the detection capabilities of Slither by integrating Interval Analysis enriched with constraints. This enhancement enabled the detection of several critical issues not natively flagged by Slither, including \defectCategory{Unreachable Code}, \defectCategory{Insufficient Parameter Validation}, and \defectCategory{Locked Ether}. A primary challenge of this approach remains the considerable effort required to formally model the complete semantics of the Solidity language. While we previously evaluated the full scope of vulnerabilities detectable via Interval Analysis, the current work extends our methodology to new categories, such as \defectCategory{Access Control}. To achieve this, we leverage our existing solution for Interval Analysis and Constraint Gathering to compute rich contextual data, which serves to improve the performance of Large Language Models (LLMs) in identifying issues and then generating scenarios in the form of tests. These tests can then be validated using either Forge or Kontrol, depending on whether symbolic or concrete execution is more appropriate for that specific defect type.

The work presented in~\cite{schiffl2021towards} addresses the challenge of establishing correctness, particularly for access control aspects, in smart contracts. The authors undertake a case study using an existing smart contract application called Palinodia, which focuses on decentralized identity management and access control for binary file integrity. Their approach involves formally specifying and verifying correctness at both the single-function level and for temporal properties of the overall application, demonstrating how verified low-level properties can support higher-level correctness. They developed natural language specifications for Palinodia's functions and translated them for verification using the \keyTerm{solc-verify}~\cite{hajdu2019solc} tool, finding it straightforward and fast for single functions. One flaw of current analysis tools that the authors point out is the lack of support for application-specific properties and verifying interacting contracts. However, their study took place before the popularization of LLMs. We believe that AI models have the potential to assist in automatically generating contract-specific properties if given enough context.

Many other notable solutions that leverage LLMs to assist in traditional static analysis approaches have recently emerged. One such notable solution is \keyTerm{IRIS}~\cite{li2025iris}, which integrates LLMs with traditional static analysis tools like \keyTerm{CodeQL} to overcome the challenges of manual taint specification crafting. A taint specification defines security-relevant data flows by identifying \keyTerm{data sources} (untrusted input), \keyTerm{sinks} (vulnerable functions), and \keyTerm{sanitizers} (functions that make data safe). IRIS leverages an LLM to automatically infer these specifications and then uses a tool like \keyTerm{CodeQL} to find vulnerable program paths. Finally, it uses the LLM again as a post-analysis filter to reduce false alarms by asking it to validate the findings. While both our approach and IRIS use an LLM to reason about the code, our methodologies differ significantly. IRIS uses the LLM to generate configurations for a traditional static analyzer and to filter its results, whereas our pipeline uses the LLM to generate executable tests that are then formally or empirically verified by a symbolic or concrete execution engine like Kontrol or Forge.

In the area of Smart Contracts vulnerability detection, \keyTerm{LLM-SmartAudit}~\cite{wei2024llm} is an innovative framework that leverages a multi-agent conversational approach with LLMs. Designed to overcome the limitations of traditional tools, its core methodology is built upon role specialization, deploying specialized agents such as a \say{Project Manager}, \say{Auditor}, and \say{Solidity Programming Expert}. These agents engage in structured conversations to analyze code and identify weaknesses, enhancing factual accuracy. While both our work and \keyTerm{LLM-SmartAudit} use LLMs to reason about Solidity code, the application and architecture are fundamentally different. \keyTerm{LLM-SmartAudit} relies on a conversational, multi-agent system to directly identify vulnerabilities, whereas our pipeline uses a single LLM for a targeted code generation task: populating test templates. In our approach, the final verdict of whether a defect exists comes from an external execution engine, not the LLM itself.

\section{Case Studies}
Initially, we identified several scenarios that are not adequately handled by either traditional analysis tools or LLMs. AI models and analysis tools that do not leverage symbolic or concrete execution appear to struggle with accurately identifying or ruling out more "dynamic" issues. These fault categories involve the \keyTerm{gas} system, user balances, access policies, or the contract's intended logical behavior. Consequently, we decided to employ either symbolic or concrete execution, depending on the issue type. We selected \keyTerm{Forge} for the concrete execution approach and \keyTerm{Kontrol} for the symbolic execution approach. The drawback of using these tools lies in their requirement for manual interaction to write tests. We propose that LLMs can perform this task if provided with sufficient context and a test template. Thus, we designed a testing methodology and template for each of the issue types we initially identified, generating tests for them using LLMs. We used \keyTerm{Google Gemini 2.5 Pro} via its official chat application for this purpose. Currently, we have experimented with \textbf{seven} defect types: \defectCategory{Reentrancy}, \defectCategory{Insufficient parameter validation}, \defectCategory{Block Environment Dependency}, \defectCategory{Faulty Assert or Reverts}, \defectCategory{Gas Costly Pattern}, \defectCategory{Access Control}, and \defectCategory{Possible Division by Zero}.

Each of the following subsections is dedicated to a specific defect. We first provide a brief definition, followed by a representative smart contract that either contains the defect or illustrates a scenario prone to false alarms from traditional tools. Finally, we present the corresponding Kontrol/Forge test, which we generated using \keyTerm{Gemini 2.5 Pro} based on our test templates.

Before introducing our examples, we must establish basic concepts regarding Kontrol and Forge:
\begin{itemize}
\item Forge fuzzes test parameters when running tests; the number of fuzz runs is customizable. Additional variables with arbitrary values can also be defined within the test method.
\item Kontrol replaces test parameters with symbolic values. Other symbolic variables can also be defined within the test method.
\item The method \codeElement{vm.deal} adds a specified amount of currency to an address's balance.
\item The function \codeElement{vm.prank} ensures all subsequent calls are performed by the designated address.
\item When using \codeElement{vm.expectRevert}, we assert that the next executed method will throw an exception. If this does not occur, the statement will throw an exception, causing the test to fail.
\item Using \codeElement{vm.assume} enforces a constraint. Forge discards fuzzing attempts for which the condition is not valid. When using Kontrol, condition will be attached to the existing condition path.
\end{itemize}

\label{sec:case_study}
\subsection{Reentrancy -- False Positive}
\defectCategory{Reentrancy} is a critical vulnerability where an external call within a function allows an attacker's contract to repeatedly re-enter that function before its state has been updated. This exploit can often lead to the complete draining of the contract's funds. This section examines a scenario that appears vulnerable but is actually a false positive.

For our first case study, we highlight an example of a contract that might seem vulnerable to \defectCategory{Reentrancy} at first glance, but in reality, is not. The code is shown in Listing~\ref{lst:reentrancy_false_alarm}. As mentioned in Section~\ref{sec:introduction}, due to its use of the \codeElement{transfer} primitive instead of \codeElement{call}, the function that sends Ether is not reentrant.

Our process for proving that the \codeElement{withdraw} function is not vulnerable to \defectCategory{Reentrancy} involves two steps. The test contract implemented to prove this aspect is displayed in Listing~\ref{lst:reentrancy_proof}. We first need to prove that the function being tested works in usual scenarios. To prove this aspect, we implement the \codeElement{test\_proofWithdrawUsuallyWorks} function. This function validates that a normal flow—where an address deposits Ether and withdraws it—behaves as expected if no malicious actions are employed. For this "positive" scenario, we use the test contract to call the methods for depositing and withdrawing Ether. For the second step, we implement a malicious actor that will attempt to perform a \defectCategory{Reentrancy} attack. The code for the attacker contract can be found in Listing~\ref{lst:reentracy_proof_attacker}. The second test method that we implemented, \codeElement{test\_proofReentrancyExploit}, is very similar to the first one. However, this time, an instance of the attacker contract calls the suspected "reentrant" method, and we expect the transaction to revert every time. Thus, we have proven that the method behaves as expected when used as intended, but will revert if the user attempts any malicious actions. If the first test passes and the second fails, then the contract is vulnerable to \defectCategory{Reentrancy}.
\begin{lstlisting}[language=Solidity, caption={Proving that the contract in Listing~\ref{lst:reentrancy_false_alarm} is not susceptible to \defectCategory{Reentrancy attacks}}, label={lst:reentrancy_proof}]
contract ReentrancySimpleTest is Test {
    
    ReentrancySimple public _contractUnderTest;
    Attacker public _attacker;
    uint256 constant ETH_UPPER_BOUND = 2 ** 45;

    function setUp() public {
        _contractUnderTest = new ReentrancySimple();
        _attacker=new Attacker(address(_contractUnderTest));
    }

    function test_proofWithdrawUsuallyWorks(uint initialDeposit) public {
        vm.assume(initialDeposit<ETH_UPPER_BOUND);
        vm.deal(address(_attacker), initialDeposit);
        vm.deal(address(_contractUnderTest), initialDeposit);
        // --- ARRANGE ---
        _contractUnderTest.deposit{value: initialDeposit}();

        // --- ACT ---
        _contractUnderTest.withdraw();
    }

    function test_proofReentrancyExploit(uint initialDeposit) public {
        vm.assume(initialDeposit<ETH_UPPER_BOUND);
        vm.deal(address(_attacker), initialDeposit);
        vm.deal(address(_contractUnderTest), initialDeposit);
        // --- ARRANGE ---
        vm.prank(address(_attacker));
        _attacker.resetAttackCount();
        _contractUnderTest.deposit{value: initialDeposit}();

        // --- ACT ---
        vm.expectRevert();
        _contractUnderTest.withdraw();
    }
    
    receive() external payable {
    }
}
\end{lstlisting}

\begin{lstlisting}[language=Solidity, caption={Attacker helper contract for the proof in Listing~\ref{lst:reentrancy_false_alarm}}, label={lst:reentracy_proof_attacker}]
contract Attacker {
    ReentrancySimple private _victim;
    uint256 public attackCallCount;
     constructor(
        address victimAddress)
    { 
        _victim = ReentrancySimple(victimAddress);
    }

    function resetAttackCount() public
    {
        attackCallCount=0;
    }

    receive() external payable {
        if (attackCallCount < 1) {
            attackCallCount++;
            _victim.withdraw();
        }
    }
}
\end{lstlisting}
\subsection{Gas Costly Pattern -- Complex Fallback}
\defectCategory{Gas Costly Pattern} defects are not traditional security flaws but rather inefficient coding practices that can result in prohibitively high transaction fees, which are measured in \keyTerm{gas}, potentially rendering a contract unusable. This section explores a specific subtype, which we term \defectCategory{Complex Fallback}, where an expensive operation in a contract's receive callback causes transactions with a limited gas supply to fail.

Listing~\ref{lst:gas_costly_pattern} contains a simple crowdfunding contract. While the business logic, contract, and its access policies are correctly implemented, an issue could gravely affect the contract's usability. This issue lies in the contract's \codeElement{receive} callback. It assigns a state variable of the contract, an operation that is expensive from a gas perspective. This can be problematic because Solidity primitives such as \codeElement{send} and \codeElement{transfer} only attach 2300 gas to the transaction when used to send Ether. Writing to the contract state consumes considerably more gas than this amount. Thus, if a contract uses one of these two functions to send currency to the example contract, the transaction will always revert. 
\begin{lstlisting}[language=Solidity, caption={A contract containing a \codeElement{receive} callback that uses too much gas}, label={lst:gas_costly_pattern}]
contract ComplexFallback {
    address private _owner;
    address private _latestDonor;

    modifier onlyOwner() {
        require(
            msg.sender == _owner,
            "Only the owner of the contract can access this"
        );
        _;
    }

    constructor() {
        _owner = msg.sender;
    }

    receive() external payable {
    //Expensive state change
        _latestDonor = msg.sender;
    }

    function withdrawFunding() external onlyOwner {
        payable(msg.sender).transfer(address(this).balance);
    }

    function getLatestDonor() external view onlyOwner returns (address) {
        return _latestDonor;
    }
}
\end{lstlisting}

We provide a possible solution to confidently prove the presence of such an issue in Listing~\ref{lst:gas_costly_pattern_proof}. Since the undesired behavior is triggered when receiving Ether, no parameters or contract state need to be considered. Consequently, using a tool such as Kontrol to generate a proof through symbolic execution is not mandatory. Using Forge to carry out concrete executions on the two test methods is sufficient to prove this defect. Each test method needs to be executed once to reveal the code fault; no fuzzing is needed. Our approach for proving the presence of this defect involves two steps (test methods).

The first method, \codeElement{test\_proveTransferWorks}, validates that sending currency to the contract works when all the gas remaining for the transaction is forwarded, thereby allowing the execution of the complex fallback. To attach this amount of gas, we use the \codeElement{call} primitive, which forwards all remaining gas by default. It also returns "true" if the transaction was successful; we validate this result through an assertion at the end. The second test, \codeElement{test\_proveTransferDoesNotWorkWithLimitedGas}, is similar. The key difference lies in the fact that this time, we use \codeElement{transfer}, and we expect the transaction to fail. The first test proves that sending currency behaves as expected when all gas is forwarded. The second test proves that the transaction reverts if only 2300 gas is forwarded. We can conclude that the \codeElement{receive} callback can only fail due to an "Out-of-Gas" exception, thus proving the \defectCategory{Complex Fallback} issue.

This testing strategy for callbacks that consume too much gas can be easily applied to any contract. The two tests remain the same; the only aspect that should be changed is the declared type and the initialization of \codeElement{\_contractUnderTest}. Because this issue is tied to a behavior that all contracts capable of receiving currency possess, no other adjustments regarding the contract's logic are necessary. An LLM can perform this data type replacement if given the template. Thus, we believe that this type of defect can be identified reliably using our approach. 
\begin{lstlisting}[language=Solidity, caption={Proof for the defect presented in Listing~\ref{lst:gas_costly_pattern}}, label={lst:gas_costly_pattern_proof}]
contract ComplexFallbackTest is Test,KontrolCheats {
    ComplexFallback public _contractUnderTest;

     function setUp() public {
        _contractUnderTest = new ComplexFallback();
    }

    function test_proveTransferWorks() public
    {
        // --- ARRANGE ---
        vm.deal(address(this), 1 ether);
        // --- ACT ---
        (bool success,)=payable(address(_contractUnderTest)).call{value: 1 ether}("");
        // --- ASSERT ---
        vm.assertTrue(success,"The transaction should work");
    }

    function test_proveTransferDoesNotWorkWithLimitedGas() public
    {
        // --- ARRANGE ---
        vm.deal(address(this), 1 ether);
        // --- ASSERT ---
        vm.expectRevert();
        // --- ACT ---
        payable(address(_contractUnderTest)).transfer(1 ether);
    }
}
\end{lstlisting}
\subsection{Insufficient Implementation \& Usage of Access Policies}
\defectCategory{Insufficient Access Policies} are defects where functions containing critical logic or actions, such as transferring Ether or destroying a contract, are not properly secured. This lack of security can permit arbitrary and potentially malicious access to sensitive operations that should be restricted.

We present a basic \defectCategory{Access Control} issue in Listing~\ref{lst:access_control_issues_example}. While we know that the \codeElement{selfdestruct} function is deprecated, it serves as an illustrative example, and this type of issue is not limited to this Solidity primitive. This defect type applies to functions that contain critical logic or actions, such as transferring Ether. Usually, these methods should be secured via access policies, preventing arbitrary, potentially malicious access. The example presented in this section implements an access modifier that only allows the address that deployed the contract to access functions to which \codeElement{onlyOwner} is attached. However, this modifier is not attached to the function containing the \codeElement{selfdestruct} primitive, thus making it accessible to any address. Because the logic and recommended access policies of a function are hard to determine via traditional static analysis, we believe that using Large Language Models might fill the gaps required to reliably prove the insufficient implementation of access policies for some functions.
\begin{lstlisting}[language=Solidity, caption={A containing an \defectCategory{Access Control} defect}, label={lst:access_control_issues_example}]
contract UnprotectedSelfdestruct {
    mapping(address => uint256) private _employees;
    address private _owner;

    modifier onlyOwner() {
        require(
            msg.sender == _owner,
            "Only the owner of the contract can access this"
        );
        _;
    }

    constructor() {
        _owner = msg.sender;
    }

    function sendSalary(address employeeAddress) external payable onlyOwner {
        _employees[employeeAddress] = msg.value;
    }

    function getSalary() external {
        require(
            _employees[msg.sender] > 0,
            "You cannot receive your salary at this moment"
        );
        _employees[msg.sender] = 0;
        payable(msg.sender).transfer(_employees[msg.sender]);
    }

    //No access modifier added, every address can call this
    function cancelContract() external {
        selfdestruct(payable(msg.sender));
    }
}
\end{lstlisting}
We highlight a test that proves this vulnerability in Listing~\ref{lst:access_control_issues_proof}. Our proof contains a single test demonstrating that an arbitrary address, not the owner, can call the function containing the \codeElement{selfdestruct} primitive. By providing the address that will call the function in the \codeElement{caller} parameter, we inform Kontrol that it should be treated as a symbolic value. We then use \codeElement{vm.assume} to attach a constraint to this symbolic value, ensuring it has a value different from the owner's. If access policies are correctly implemented, this call should revert for any user who is not the contract's owner, and the test should pass. However, for the example from Listing~\ref{lst:access_control_issues_example}, the function does not revert, thus causing the test to fail.

We can extract a template from this example. It should contain a function call made by a symbolic address, precisely as the \codeElement{caller} parameters are used. A crucial point is imposing constraints on the symbolic value that reflect the fact that the symbolic address should not be part of the group of users typically allowed to access the function. These constraints used to be hard to determine automatically. However, we hypothesize that LLMs have potential to find recommended access policies for a function.
\begin{lstlisting}[language=Solidity, caption={Proof for the defect presented in Listing~\ref{lst:access_control_issues_example}}, label={lst:access_control_issues_proof}]
contract UnprotectedSelfdestructTest is Test {
    
    UnprotectedSelfdestruct public _contractUnderTest;

    function setUp() public {
        _contractUnderTest = new UnprotectedSelfdestruct();
    }

    function test_accessControl(address caller) public {
        // --- ARRANGE ---
        vm.assume(caller!=address(this));
        vm.prank(caller);
        // --- ASSERT ---
        vm.expectRevert();
        // --- ACT ---
        _contractUnderTest.cancelContract();
    }
}
\end{lstlisting}
\section{Towards a Combined Approach}
\label{sec:our_approach}
As discussed in the preceding sections, each technique possesses inherent limitations. Large Language Models do not provide consistent responses, particularly when lacking sufficient context and examples. Fuzz testing and symbolic execution-based testing necessitate test contracts tailored specifically for the contract under analysis. Furthermore, symbolic execution can be slow in terms of analysis time. While being fast and capable of detecting an impressive number of faults, Slither is unable to correctly label certain defect categories. It also struggles with specific scenarios that highlight defects it otherwise labels correctly. These flaws do not stem from implementation issues but rather from the fundamental nature of the analysis itself, as each approach has its inherent strengths and weaknesses.

Consequently, we conjecture that combining all these approaches into a single solution will leverage their individual strengths while mitigating some of their weaknesses. Our approach is structured in three steps: collecting contract information, generating symbolic or fuzz tests, and executing the tests. We use Slither to gather useful contextual information about the contract and other relevant data regarding the defect type under investigation. This information, along with a test template, is then forwarded to the LLM for test generation. The generated tests are subsequently executed using either Forge or Kontrol. Depending on the results, the issue is either signaled or not. If the process fails before test execution and result collection, issues are not signaled or are reported with a lower confidence rate. The method for reporting error cases depends on the type of vulnerability being investigated. For example, when investigating \defectCategory{Access Control} issues, a lower confidence rate might be reported if the contract contains functions with critical operations such as \codeElement{selfdestruct} but lacks custom-built access modifiers, even if the presence of insufficient access policies cannot be definitively proven as shown in Section~\ref{sec:case_study}. We propose to design custom flows for each defect type, as we believe this is the optimal decision given that each defect type requires specific information from Slither and a distinct number of tests with their own unique testing template. It is important to note that when we prove the presence or absence of a defect, we are formally or empirically proving that the contract is vulnerable or not vulnerable to the specific attack scenarios encoded in our test templates, not all possible variations of an attack.

\paragraph{Stage 1: Contract preprocessing}
We begin our analysis process by using custom-built implementations based on Slither modules to extract specific information. Each defect type necessitates a different set of characteristics to be extracted, therefore requiring its own custom implementation. We chose Slither because its modules provide rich, structured data about the smart contract at various levels of abstraction. While we only mention a few examples—such as contract-level details like state variables, method-level information like parameters, and statement-level data like the variables in an instruction—the framework exposes far more granular information. This comprehensive access to the contract's structure allows us to efficiently check for the specific prerequisite patterns of each defect type. Thus, by analyzing the data provided by Slither, we can verify the prerequisites for each defect type through a process similar to pattern matching. For example, when investigating \defectCategory{Block Environment Dependency}, we search for methods that read block information, such as \codeElement{block.timestamp}. If any such functions are found, we return the statements where those built-in variables are used. This initial step effectively rules out scenarios where a certain issue is not present. Continuing with the \defectCategory{Block Environment Dependency} example, if no block-related built-in variables are detected in the contract's code, there is no need to continue the analysis process, as we can confidently state that this defect is not present. Similarly, for \defectCategory{Reentrancy}, we identify functions containing external calls and check if they are followed by state-modifying instructions.
\paragraph{Stage 2: Test generation using LLMs  }
After retrieving relevant information for the defect type under analysis, we proceed to the test generation step. We provide the gathered information along with a test template to the LLM. The LLM is tasked with populating this template with contract-specific information. The complexity of inserting the required information into the template varies depending on the issue type. For some issues, such as \defectCategory{Complex Fallback}, the model only needs to replace the type of the contract being tested. For other defects, more complex "reasoning" is required; examples of these more difficult tasks include inferring recommended access policies for a contract or handling intricate function calls and setups. So far, LLMs have been capable of creating tests for our examples. For some experiments, the tests generated were correct, and some required minimal adjustments. As expected, more complex generation tasks yield tests that need adjusting. 
A core aspect of our template implementation methodology is using code comments to provide direct instructions to the LLM, leveraging its powerful instruction-following capabilities. The template for the Complex Fallback defect, shown in Listing~\ref{lst:complex_fallback_template}, exemplifies this approach. While the test methods and assertions are already complete, the comments guide the LLM to generate the final, contract-specific code.

\begin{lstlisting}[language=Solidity, caption={The template used to generate the test presented in Listing~\ref{lst:gas_costly_pattern_proof}},label={lst:complex_fallback_template}]
// SPDX-License-Identifier: MIT
pragma solidity 0.8.29;
import {Test} from "../../lib/forge-std/src/Test.sol";
//Replace this import with one corresponding to the contract type being tested, the file has the same name as the contract and is located in the same folder path as the example
import {ContractUnderTest} from "../../src/ContractUnderTest.sol";

//Append the type of the contract being tested to the name of the Test contract
contract Test is Test {
 //Replace the type "ContractUnderTest" of _contractUnderTest with the type of the contract that is currently being analyzed"
        ContractUnderTest public _contractUnderTest;

    // If the constructor of the contract under test has parameters, the "setUp" function should have the same parameters 
     function setUp() public {
        //Initialize the contract being tested with the correct constructor, use the correct parameters
        
        //If the constructor is payable, use "vm.deal" to add balance to the current contract before initializing the contract under test
        
        _contractUnderTest = new ContractUnderTest();
    }

    //We will use this method to prove that the transfer works when all gas is forwarded to the transaction
    function test_proveTransferWorks() public
    {
        address sender = makeAddr("sender");
        vm.deal(sender, 1 ether);
        vm.prank(sender);
        (bool success,)=payable(address(_contractUnderTest)).call{value: 1 ether}("");
        vm.assertTrue(success,"The transaction should work");
    }

    //We will use this method to prove that the transfer fails when a hardcoded amount of 2300 gas is attached to the transaction
    function test_proveTransferDoesNotWorkWithLimitedGas() public
    {
        address sender = makeAddr("sender");
        vm.deal(sender, 1 ether);
        vm.prank(sender);
        vm.expectRevert();
        payable(address(_contractUnderTest)).transfer(1 ether);
    }
}
\end{lstlisting}

\paragraph{Stage 3: Test execution}
After obtaining the tests, we integrate them into a pre-configured Forge/Kontrol environment. We then execute the tests and retrieve the success or failure result for each test method. The number of tests and the interpretation of their results vary depending on the issue type. For instance, as presented in Section~\ref{sec:case_study}, \defectCategory{Reentrancy} requires two tests, while \defectCategory{Access Control} requires only one. Furthermore, in our example, if both tests pass for \defectCategory{Reentrancy}, the contract is not vulnerable. The opposite is true for \defectCategory{Access Control} and \defectCategory{Complex Fallback}: if their respective tests pass, then the issue is present. If the process fails during this stage, we may provide lower confidence detections or no detections at all.

While we have not yet implemented a complete practical solution for our proposed workflow, our previous experiments and existing implementations enable us to evaluate its feasibility and implementation complexity. We have already experimented with extending Slither for new analysis types and use cases~\cite{from2023,from2024}, and we plan to implement the preprocessing stage soon. Furthermore, our prior experience with programmatically interacting with LLMs~\cite{SAS2025} will significantly accelerate the development of the second stage. Kontrol and Forge are the newest additions to our toolbox; however, we have already manually executed meaningful scenarios that demonstrate their potential.

\section{The Proposed Approach at Work on an Example}
\label{sec:work-ex}
To illustrate our pipeline's workflow, this section details the analysis of the contract from Listing~\ref{lst:gas_costly_pattern} for a Complex Fallback vulnerability. The process is described step-by-step, corresponding to the three stages outlined in Section~\ref{sec:our_approach}.

\paragraph{Stage 1}
In the first stage, our pipeline checks the primary prerequisite for a \defectCategory{ComplexFallback} defect: the contract's ability to receive Ether through a simple transfer. The vulnerability specifically arises when a contract's receive() function, which is automatically executed upon a direct Ether transfer, is too gas-intensive. Therefore, our Slither-based tool scans the contract from Listing~\ref{lst:gas_costly_pattern} to confirm a receive() function is implemented. Since the function is present, the contract is a candidate for this vulnerability, and the analysis proceeds. If a receive() function were absent, the contract could not be affected by this specific defect, and the analysis would terminate here with a negative finding.

\paragraph{Stage 2}
In the second stage, we use the template from Listing~\ref{lst:complex_fallback_template} to generate the final test. The pipeline sends this template to an LLM, such as Gemini 2.5 Pro, with instructions to populate it using the contract's specific details. For this example, the primary instruction is to replace the placeholder ContractUnderTest with ComplexFallback. The LLM then returns a complete and compilable Forge test file, ready for execution. If this generation step fails, the analysis is terminated, and the contract is flagged as potentially vulnerable, albeit with a lower confidence rating.

\paragraph{Stage 3}
In the final stage, the generated test is executed using Forge. We chose Forge for this defect because the analysis only requires two concrete executions, not the complex path exploration for which Kontrol is designed. The pipeline runs the test and interprets the results:
\begin{itemize}
    \item \codeElement{test\_proveTransferWorks()} is expected to pass, confirming that the contract can receive Ether when sufficient gas is provided via \codeElement{call()}.
    \item \codeElement{test\_proveTransferDoesNotWorkWithLimitedGas()} is also expected to pass, as it correctly anticipates that the \codeElement{transfer()} call (with its 2300 gas limit) will revert.
\end{itemize}

\noindent Since both tests pass, our pipeline confidently flags the contract as vulnerable to the Complex Fallback defect. This confirms that the contract's receive() function is too gas-intensive for simple Ether transfers.

\section{Challenges \& Limitations}
\label{sec:chal-and-lim}
\paragraph{The consistency of LLMs}
Since LLMs are at the core of our implementation, many of our solution's challenges and limitations are inherently tied to their performance in test generation and the difficulty of obtaining consistent responses. Obtaining consistent and structured output from LLMs has been a challenge since their integration into various software solutions a few years ago. While some model developers, such as OpenAI, now offer options to define output structure in requests~\cite{OpenAI_Structured_Outputs}, our experiments showed that this feature is still not mature, despite providing a noticeable improvement in structural consistency. Regarding response content consistency, we have mitigated this issue by providing test templates and contextual information. Although the generated test code was not always identical, variations were limited to aspects that did not affect correctness, such as test method parameter names, local variable names, and assertion messages. We did not experiment much with top-tier reasoning models such as \keyTerm{OpenAI O3}~\cite{OpenAI_Reasoning_Models}, but we expect them to handle more complex scenarios compared to regular models.

\paragraph{The cost of using LLMs}
Another limitation of our solution is how software systems interact with models. Most state-of-the-art models are proprietary and cannot be deployed locally. Instead, they are accessed via APIs offered by developers, which can raise cost concerns depending on the model and the number of tokens used. An alternative is open-source models like the DeepSeek models~\cite{DeepSeek} and Llama models~\cite{Llama}. Even though these models are publicly available and can be deployed locally, they still demand considerable computational power. A traditional consumer system is usually not nearly enough to run their larger variants. While the smaller variants offer satisfactory performance in general-purpose tasks, there is still a noticeable gap when performing specialized tasks like ours.

\paragraph{Forge \& Kontrol integration}
Another challenge lies in reliably integrating Forge and Kontrol into our solution. Our use case would ideally require an \say{API mode} for these tools. Even though this feature is not currently available, both tools are open source, so it could be implemented, though it would demand additional development time. While automatically using them from the command line would yield results, capturing the output via string matching might not be consistent, and it is not our preferred approach. Currently, we have been able to overcome this limitation by using an LLM to convert the unstructured CLI output into a structured format while also stripping away details that do not aid our process. However, having a direct integration with Kontrol and Forge via an API would still be the preferred approach due to its considerably greater robustness. 
\section{Conclusions}
\label{sec:conclusions}
In this paper, we propose a combined approach aimed at accurately classifying code faults previously misidentified by state-of-the-art tools or Large Language Models (LLMs) independently. While still in its incipient phase, we hypothesize that this approach holds significant potential to enhance the accuracy of code defect detection in Smart Contracts, particularly for issues currently undetected in most cases. Our approach follows a three-step workflow that integrates the strengths of Slither, LLMs, and Forge/Kontrol.

We manually conducted experiments across seven defect categories to evaluate the potential of LLMs in generating specialized Smart Contract tests. Our results demonstrate that AI models are capable of generating tests that accurately determine a contract's vulnerability status. We presented three such scenarios in greater detail, which were initially misclassified by traditional analysis tools or LLMs. Through our experiments, we reliably proved the presence of specific defects in these cases, and their absence for the tested attack scenarios.

For future work, we aim to develop a practical implementation of this concept. Our prior experience in building custom detectors based on Slither will be invaluable for implementing the first step of our approach. Subsequently, we intend to experiment with additional vulnerability types and design corresponding test templates. Regarding LLM interaction, we will explore various prompting techniques and model variants to identify those yielding the most optimal and consistent results for our use case. Moving to the next step, we will investigate methods for programmatic interaction with Forge/Kontrol. If a direct API is unavailable, our fallback option will be command-line integration. Should this prove unreliable, we will consider extending these tools by implementing the necessary features. Regardless of the integration method, we perceive substantial potential in automatically generating input for testing tools, viewing it as a highly powerful approach from a detection capability perspective.

\subsubsection*{Acknowledgements}
Part of this work was supported by a grant from the Romanian Ministry of Research, Innovation and Digitization, CNCS/CCCDI - UEFISCDI, project 86/2025 ERANET-CHISTERA-IV-SCEAL, within PNCDI IV.

\nocite{*}
\bibliographystyle{eptcs}
\bibliography{bibliography}
\end{document}